\documentclass[aps,nofootinbib,floatfix,amsmath,amssymb,preprint,groupedaddress,groupedaddress,prd]{revtex4-1}

\usepackage{hyperref}

 \usepackage{graphicx}
 
\usepackage[font=small,labelfont=bf]{caption}
\usepackage[english]{babel}

\usepackage{xspace}

\usepackage{bm}

\newcommand{\mxnewcommand}[2]{\newcommand{#1}{\ensuremath{#2}\xspace}}

\newcommand{\xnewcommand}[2]{\newcommand{#1}{#2\xspace}}

\mxnewcommand{\gev}{\,\text{GeV}}
\mxnewcommand{\tev}{\,\text{TeV}}
\mxnewcommand{\invfb}{/\text{fb}}

\newcommand{\roots}[1]{\ensuremath{\sqrt{s}={#1}\tev}}

\mxnewcommand{\mhalf}{m_{1/2}}
\mxnewcommand{\mzero}{m_0}
\mxnewcommand{\azero}{A_0}
\mxnewcommand{\tanb}{\tan\beta}
\mxnewcommand{\sgnmu}{\text{sign}\,\mu}
\mxnewcommand{\sgnb}{\text{sign}\,b}
\mxnewcommand{\msbar}{\overline{\text{MS}}}
\mxnewcommand{\invalpha}{1/\alpha_{\text{em}}(M_Z)^{\msbar}}
\mxnewcommand{\alphas}{\alpha_s(M_Z)^{\msbar}}
\mxnewcommand{\mt}{m_t^\text{Pole}}
\mxnewcommand{\mb}{m_b(m_b)^{\msbar} }
\mxnewcommand{\mz}{M_Z}
\mxnewcommand{\mhu}{m_{H_u}}
\mxnewcommand{\mhd}{m_{H_d}}
\mxnewcommand{\tanbsq}{\tan^2\beta}
\mxnewcommand{\mplanck}{M_\text{P}}
\mxnewcommand{\msusy}{M_\text{SUSY}}
\mxnewcommand{\mgut}{M_\text{GUT}}
\mxnewcommand{\msoft}{m_\text{soft}}
\mxnewcommand{\mts}{m_t}
\mxnewcommand{\mbs}{m_b}
\mxnewcommand{\invalphas}{1/\alpha_{\text{em}} }
\mxnewcommand{\alphass}{\alpha_s}

\renewcommand{\(}{\left(}
\renewcommand{\)}{\right)}
\renewcommand{\[}{\left[}
\renewcommand{\]}{\right]}

\mxnewcommand{\like}{\mathcal{L}}
\mxnewcommand{\prior}{\pi}
\mxnewcommand{\params}{m}
\mxnewcommand{\ev}{\mathcal{Z}}
\mxnewcommand{\point}{\vec{x}}
\mxnewcommand{\model}{\text{model}}
\mxnewcommand{\data}{\text{data}}
\mxnewcommand{\pd}{\,\prod\text{d}}

\newcommand{\priorf}[1]{\ensuremath{\prior(#1)}}
\newcommand{\likef}[1]{\ensuremath{\like(#1)}}

\newcommand{\pg}[2]{\ensuremath{p(#1\,\bm{|}\,#2)}}
\newcommand{\p}[1]{\ensuremath{p(#1)}}

\newcommand{\dpartial}[2]{\ensuremath{\frac{\partial #1}{\partial #2}}}
\newcommand{\indpartial}[2]{\ensuremath{{\partial #1}/{\partial #2}}}

\newcommand{\s}[1]{\ensuremath{\tilde{#1}}}
\newcommand{\neut}[1]{\ensuremath{{\chi}^0_{#1}}}

\newcommand{\ms}[1]{\ensuremath{m_{\s{#1}}}}
\newcommand{\mneut}[1]{\ensuremath{m_{\neut{#1}}}}

\mxnewcommand{\sigsip}{\sigma^{\text{SI}}_p}
\mxnewcommand{\abund}{\Omega h^2}
\newcommand{\br}[1]{\ensuremath{\text{BR}}(#1)}
\mxnewcommand{\bsg}{\br{B_s\to X_s\gamma}}
\mxnewcommand{\bsmm}{\br{B_s\to \mu\mu}}
\mxnewcommand{\btn}{\br{B_u\to \tau\nu}/\br{B_u\to \tau\nu}|_{\text{SM}}}
\mxnewcommand{\damu}{\delta a_\mu}
\mxnewcommand{\mh}{m_h}
\mxnewcommand{\ma}{m_A}
\mxnewcommand{\mw}{M_W}
\mxnewcommand{\dmbs}{\Delta M_{B_s}}
\mxnewcommand{\sineff}{\sin^{2} \theta_{\ell,\text{eff}}}

\mxnewcommand{\pmm}{(\mzero,\,\mhalf)}
\mxnewcommand{\pat}{(\azero,\,\tanb)}
\mxnewcommand{\pcs}{(\mneut{1},\,\sigsip)}
\xnewcommand{\stauc}{stau-coannihilation}
\xnewcommand{\Stauc}{Stau-coannihilation}

\let\oldcite\cite
\renewcommand{\cite}{~\oldcite}
\newcommand{\reftable}[1]{Table~\ref{#1}} 
\newcommand{\reffig}[1]{Fig.~\ref{#1}}       
 
\newcommand{\refeq}[1]{Eq.~(\ref{#1})}
\newcommand{\refsec}[1]{Sec.~\ref{#1}}
\newcommand{\refcite}[1]{Ref.\cite{#1}}            
\xnewcommand{\eg}{\textit{e.g.,}}
\xnewcommand{\ie}{\textit{i.e.,}}
\xnewcommand{\latincf}{\textit{c.f.,}}
\mxnewcommand{\dash}{\text{, }}
\newcommand{\ic}[1]{``#1''}

\newcommand{\beq}{\begin{equation}}
 \newcommand{\eeq}{ \end{equation}}
 \newcommand{\pn}[2]{\ensuremath{#1\cdot10^{#2}}} 

\begin{document}

\title{CMSSM, naturalness and the \ic{fine-tuning price} of the Very Large Hadron Collider}
  
\author{Andrew Fowlie}
\email{Andrew.Fowlie@KBFI.ee}
\affiliation{National Institute of Chemical Physics and Biophysics, Ravala 10,
Tallinn 10143, Estonia}

\date{\today}

%%%%%%%%%%%%%%%%%%%%%%%%%%%%%%%%%%%%%%%%%%%%%%%%%%%%%%%%%%%%%%%%%%%%%%%%%%
\begin{abstract}
The absence of supersymmetry or other new physics at the Large Hadron Collider (LHC) has lead many to question naturalness arguments. With Bayesian statistics, we argue that natural models are most probable and that naturalness is not merely an aesthetic principle. We calculate a probabilistic measure of naturalness, the Bayesian evidence,  for the Standard Model (SM) with and without quadratic divergences, confirming that the SM with quadratic divergences is improbable. We calculate the Bayesian evidence for the Constrained Minimal Supersymmetric Standard Model (CMSSM) with naturalness priors in three cases: with only the \mz measurement; 
with the \mz measurement and LHC measurements; and with the \mz measurement, \mh measurement and a hypothetical null result from a \roots{100} Very Large Hadron Collider (VLHC) with $3000\invfb$. 
The \ic{fine-tuning price} of the VLHC given LHC results would be $\sim400$, which is slightly less than that of the LHC results given the electroweak scale ($\sim500$).
\end{abstract}
%%%%%%%%%%%%%%%%%%%%%%%%%%%%%%%%%%%%%%%%%%%%%%%%%%%%%%%%%%%%%%%%%%%%%%%%%%

\maketitle

\section{\label{Section:Introduction}Introduction}
Weak-scale supersymmetry (SUSY)\cite{Salam:1974yz,Haber:1984rc,Nilles:1983ge,Martin:1997ns} was supposed to solve the naturalness problem of the Standard Model (SM)\cite{Gildener:1976ai,Susskind:1978ms}, but it was absent in the ATLAS\cite{ATLAS-CONF-2013-047} and CMS\cite{Chatrchyan:2014lfa} searches at the Large Hadron Collider (LHC) in $20\invfb$ with center-of-mass energies of \roots{7} and \roots{8}. Although ATLAS and CMS will continue their searches for SUSY at \roots{13}, a new \roots{100} Very Large Hadron Collider (VLHC) might be built\cite{FCC}.

There are numerous motivations for SUSY. The theoretical motivations for SUSY (see \eg \refcite{Dine:2007zp}) are, \textit{inter alia,} that it completes the maximal symmetries of the $S$-matrix and connects with gravity and superstrings. The phenomenological and experimental motivations for SUSY (see \eg \refcite{Baer:2006rs}) are that it unifies the gauge couplings at the anticipated scale, that the lightest SUSY particle could explain the measured abundance of dark matter in the Universe and that it predicts that the mass of the lightest Higgs boson is $\mh\lesssim135\gev$. Perhaps the strongest motivation for SUSY, however, is that it solves the technical naturalness problem of the SM, if SUSY particles are sufficiently light. The LHC results, however, suggest that SUSY particles might not be sufficiently light\cite{Giudice:2013yca,Feng:2013pwa} and have lead many to question naturalness arguments\cite{Foot:2013hna,Dubovsky:2013ira,Heikinheimo:2013fta,Farina:2013mla,deGouvea:2014xba}.

We argue in \refsec{Section:Evidence} that the best measure of naturalness is Bayesian evidence and measure naturalness in the SM in \refsec{Section:SM_evidence} and  in the Constrained Minimal Supersymmetric SM (CMSSM)\cite{Chamseddine:1982jx,Arnowitt:1992aq,Kane:1993td} in \refsec{Section:CMSSM_evidence} by calculating their Bayesian evidences with \ic{honest} or \ic{naturalness} priors. We evaluate the consequences for naturalness of hypothetical null results from a \roots{100} VLHC with Bayesian statistics, \ie the \ic{fine-tuning price} of the VLHC\cite{Chankowski:1997zh,Barbieri:1998uv,Strumia:2011dv}, by calculating the Bayesian evidence in this scenario. Learning this \ic{price} could motivate building the VLHC\cite{nima}. We argue that our comparison between the SM and the CMSSM was fair in \refsec{Section:Fair_Comp}. We discuss the $\mu$-problem of the MSSM\cite{Kim:1983dt} in the context of Bayesian statistics in \refsec{Section:Mu_problem}, and conclude in \refsec{Section:Conclusions}.  For similar analyses, see \eg \refcite{Giusti:1998gz,Allanach:2006jc,Allanach:2007qk,Feroz:2008wr,Cabrera:2012vu,Kim:2013uxa,Balazs:2013qva}.

\section{\label{Section:Evidence}Bayesian evidence}
For a pedagogical introduction to Bayesian statistics, see \eg \refcite{Trotta:2008qt}.  In Bayesian statistics, probability is a numerical measure of belief in a proposition. With Bayes' theorem, our belief in a model given experimental data is given by 
\beq
\label{Eqn:Bayes_Theorem}
\pg{\model}{\data} = \frac{\pg{\data}{\model} \times \p{\model}}{\p{\data}},
\eeq
where $\ev\equiv\pg{\data}{\model}$ is the \textit{Bayesian evidence,} $\p{\model}$ is our prior belief in the model, and $\p{\data}$ is a normalization constant.
We can eliminate the normalization constant if we consider a ratio of probabilities for $\model_a$ and $\model_b$;
\beq
\label{Eqn:Bayes_Factor}
\underbrace{\frac{\pg{\model_a}{\data} }{ \pg{\model_b}{\data}}}_{\text{Posterior odds, }\theta^\prime} = \underbrace{\frac{\pg{\data}{\model_a} }{ \pg{\data}{\model_b}}}_{\text{Bayes-factor, }B} \times \underbrace{\frac{ \p{\model_a} }{\p{\model_b}}}_{\text{Prior odds, }\theta}.\\
\eeq
Our \textit{prior odds}, $\theta$, is a numerical measure of our relative belief in $\model_a$ over $\model_b$, \textit{before} considering experimental data. 
The Bayes-factor, $B$, updates our \textit{prior odds,} $\theta$, with the experimental data, resulting in our \textit{posterior odds,} $\theta^\prime$. Our \textit{posterior odds}  is a numerical measure of our relative belief in $\model_a$ over $\model_b$, \textit{after} considering experimental data. The Bayes-factor is the ratio of the models' evidences. 

Let us make our discussion more concrete. From an experiment, one can construct a ``likelihood function'' giving the frequentist probability of obtaining the data, given a particular point, \point, in a model's parameter space,
\beq
\likef{\point} = \pg{\data}{\point,\model}.
\eeq
The likelihood function for a measurement is typically a Gaussian function (by the central limit theorem). It could be, \eg the probability of measuring a Higgs mass $\mh=125\gev$ given a particular parameter point \point in a SUSY model. With Bayes' theorem, it can be readily  shown that the evidence is an integral over the likelihood,
\beq
\label{Eqn:Evidence}
\ev = \int \likef{\point} \priorf{\point} \pd x,
\eeq
where $\priorf{\point}\equiv\pg{\point}{\model}$ is our \textit{prior}; our prior belief in the model's parameter space. Priors are somewhat subjective and there might exist a spectrum of assigned priors amongst investigators. All investigators, however, will make identical conclusions from the evidence, if the likelihood is sufficiently informative. 

Because individual evidences are somewhat meaningless (\eg the evidence has dimension $[1/\data]$), it is necessary to compare an evidence against that of a reference model with a Bayes-factor. If the Bayes-factor is greater than (less than) one, the model in the numerator (denominator) is favored. The interpretation of Bayes-factors is somewhat subjective, though we have chosen the Jeffreys' scale, \reftable{Table:Jefferys_Scale}, to ascribe qualitative meanings to Bayes-factors.  If a Bayes-factor is sufficiently large, all investigators will conclude that a particular model is favorable, regardless of their prior odds for the models. The Jeffreys' scale is, however, only a guide for interpreting a Bayes-factor; the full result is the posterior odds found by multiplying the Bayes-factor by the prior odds in \refeq{Eqn:Bayes_Factor}.

\begin{table}[ht]
\centering
\begin{ruledtabular}
\begin{tabular}{llll}
Grade   & Bayes-factor, $B$             & Preference for model in numerator\\
\hline
0       & $B\le1$                                       &   Negative\\
1       & $1<B\le3$                               &  Barely worth mentioning\\
2       & $3<B\le20$                            &  Positive\\
3       & $20<B\le150$                       &  Strong\\
4       & $B>150$                                 &  Very strong\\
\end{tabular}
\end{ruledtabular}
\caption{The Jeffreys' scale for interpreting Bayes-factors\cite{nla.cat-vn759870}, which are ratios of evidences. We assume that the favored model is in the numerator, though this could be readily inverted.}
\label{Table:Jefferys_Scale}
\end{table}

\begin{figure}[ht]
\centering
\includegraphics[height=0.49\linewidth]{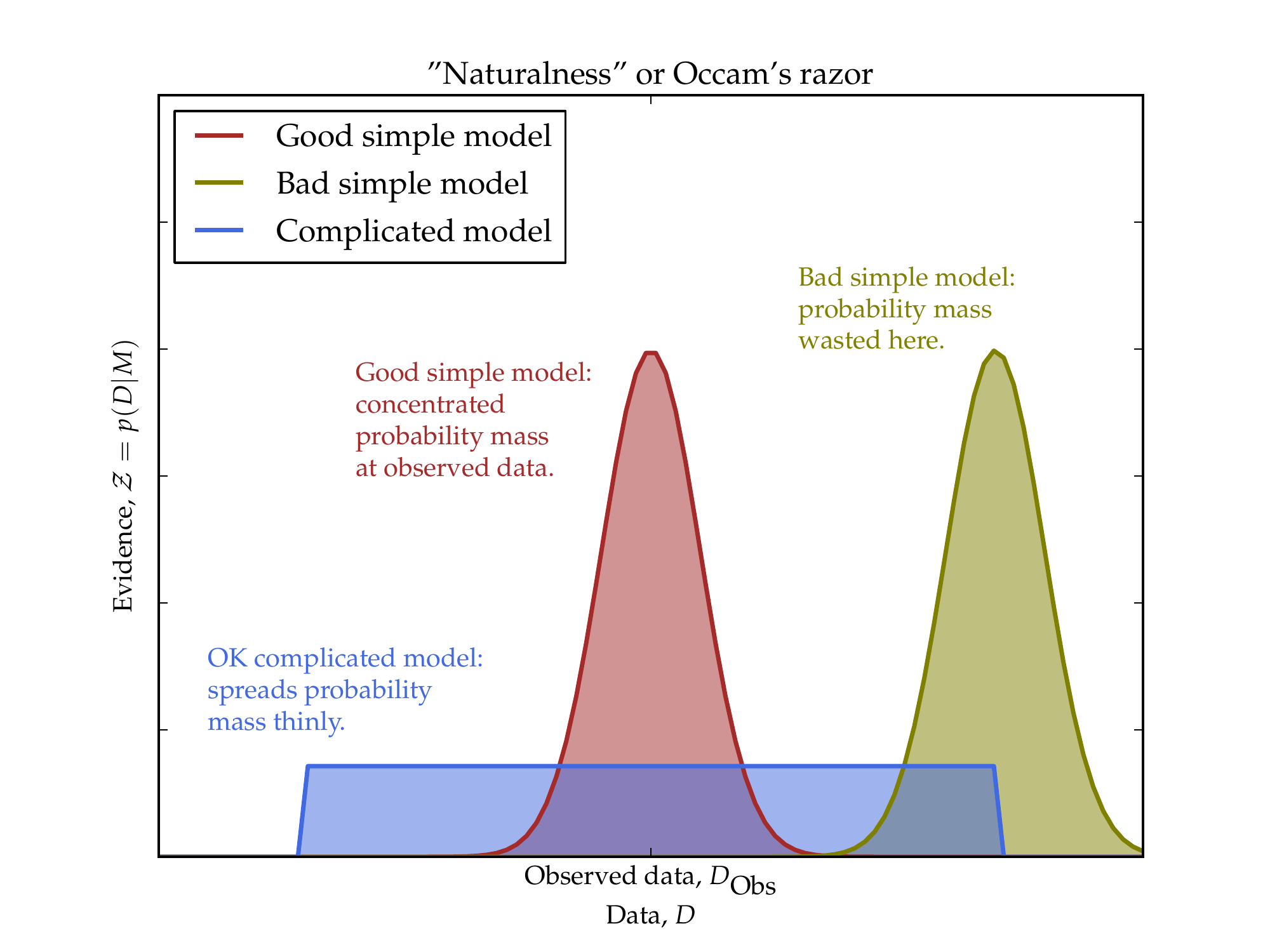}
\caption{Illustration of the evidence, interpreted as a sampling distribution, originally from \refcite{MacKay91}. The observed evidence is the evidence evaluated at the observed data. The red line shows a model that concentrates its probability mass at the observed data: it is a good, simple model. The green line shows a model that concentrates its probability mass away from the observed data: it is a bad, simple model. The blue line shows a model that thinly spreads its probability mass around the observed data: it is an OK, complicated model. 
}
\label{Fig:EvidenceExample}
\end{figure}

The Bayes-factor quantitatively incorporates a principle of economy widely-known as Occam's razor and in physics as \ic{fine-tuning} or \ic{naturalness}\cite{2009arXiv0903.4055G,Barbieri:1987fn,Ellis:1986yg}. It is insightful to consider the evidence $\ev = \pg{\data}{\model}$ a function of the data normalized to unity, \ie as a sampling distribution\cite{MacKay91}.  Natural models \ic{spend} their probability mass near the obtained  data, \ie a large fraction of their parameter space agrees with the data. Complicated models squander their probability mass away from the obtained data. This is illustrated in \reffig{Fig:EvidenceExample}. Bayesian statistics formalizes Occam's razor, fine-tuning and naturalness arguments. Naturalness is no longer a nebulous, aesthetic criterion; it is formalized and justified by Bayesian statistics.

We measure the \ic{fine-tuning price} of new experimental data with a partial Bayes-factor. A partial Bayes-factor, $P$, updates our relative belief in $\model_a$ over $\model_b$ with new experimental data,
\beq
P \cdot \frac{\pg{\model_a}{\data} }{ \pg{\model_b}{\data}} = \frac{\pg{\model_a}{\data + \text{new \data}} }{ \pg{\model_b}{\data + \text{new \data}}}.
\eeq
It can be readily shown that a partial Bayes-factor is a ratio of Bayes-factors,
\beq
P =  \frac{\pg{\data + \text{new \data}}{\model_a} }{ \pg{\data + \text{new \data}}{\model_b}}  \frac{ \pg{\data}{\model_b}}{\pg{\data}{\model_a} }.
\eeq
See \eg \refcite{Balazs:2013qva} for a comprehensive discussion of partial Bayes-factors. Having introduced our formalism, we are ready to calculate evidences in the SM and CMSSM.

\section{\label{Section:SM_evidence} Bayesian evidence for the Standard Model}
If the SM is coupled to the Planck scale,  it suffers from a well-known fine-tuning problem, the \ic{hierarchy problem}\cite{Gildener:1976ai,Susskind:1978ms}. The dimension-two coupling, $\mu^2$, in the Higgs potential,
\beq
V = \mu^2 \phi^2 + \lambda \phi^4,
\eeq
must be incredibly fine-tuned.  The dressed coupling must be $\sim-(100\gev)^2$, but the bare coupling receives a positive quadratic correction $\sim\mplanck^2$. Let us calculate the evidence for the SM, given the electroweak scale and that the Higgs mass is $\sim125\gev$.  Whilst naively our Higgs potential is described by $\mu^2$ and $\lambda$, let us instead write the dressed dimension-two coupling as the sum of a bare coupling and a quadratic correction,
\beq
\mu^2 = \mu_0^2 + \Delta \mu^2,
\eeq
and treat $\mu_0^2$, $\Delta \mu^2$ and $\lambda$ as separate parameters. \textit{A priori,} if the SM is coupled to the Planck scale, \mplanck, we expect that $\Delta \mu^2 \sim \mplanck^2$, and that $\lambda\sim1$, whereas we have no idea about the scale of $\mu_0^2$. Let us formalize these thoughts with logarithmic, scale invariant priors $\priorf{x}\propto 1/x$;
\begin{align}
\Delta \mu^2 &\text{ between } 10^{36} \text{ and } 10^{40} \gev^2,\\
\mu_0^2 &\text{ between } 10^{0} \text{ and } 10^{40} \gev^2,\\
\lambda &\text{ between } 10^{-3} \text{ and } 10^{1} .
\end{align}
We also note that \textit{a priori} $\mu_0^2$ could be positive or negative.

We calculate the evidence for the SM given the \mz measurement\cite{Beringer:1900zz} and the LHC $\mh\sim125\gev$ measurement\cite{Beringer:1900zz,Chatrchyan:2012ufa,Aad:2012tfa}. We approximate the likelihood functions for the measurements of \mz and \mh as Dirac delta functions;
\begin{align}
 \label{Eqn:SM_Evidence}
\ev_{\text{only }\mz} &= \frac { 
\int \delta(\mz - 91.1876\gev)\frac{\text{d}\mu_0^2}{\mu_0^2}\frac{\text{d}\lambda}{\lambda}\frac{\text{d}\Delta\mu^2}{\Delta\mu^2}
}{
\int \frac{\text{d}\mu_0^2}{\mu_0^2}\frac{\text{d}\lambda}{\lambda}\frac{\text{d}\Delta\mu^2}{\Delta\mu^2}
},\\\nonumber
\ev_{\mh\text{ and }\mz}  &= \frac { 
\int \delta(\mz - 91.1876\gev) \delta(\mh - 125.9\gev) \,\frac{\text{d}\mu_0^2}{\mu_0^2}\frac{\text{d}\lambda}{\lambda}\frac{\text{d}\Delta\mu^2}{\Delta\mu^2}
}{
\int \frac{\text{d}\mu_0^2}{\mu_0^2}\frac{\text{d}\lambda}{\lambda}\frac{\text{d}\Delta\mu^2}{\Delta\mu^2}
}.
\end{align}
The denominators normalize our logarithmic priors. 
We integrate the Dirac delta functions with tree-level formulas for the Higgs and $Z$-boson masses (see \eg \refcite{Cheng:1985bj}),
\begin{align}
 \mh &= \sqrt{-2\mu^2},\\
 \mz &= g \sqrt{\frac{-\mu^2}{2\lambda}}.
\end{align}
We calculate evidences by performing the integrals in \refeq{Eqn:SM_Evidence} for two models:
\begin{enumerate}
\item The SM with quadratic divergences, $\Delta\mu^2\sim\mplanck^2$ , and
\item The SM without quadratic divergences,  $\Delta\mu^2=0$. \refcite{Bardeen:1995kv,Meissner:2006zh} argue that quadratic divergences vanish in theories with classical scale invariance without modifications to the $Z$-boson or Higgs boson masses.  
\end{enumerate}

The resulting evidences are in  \reftable{Table:SM_CMSSM_evidences}. Unsurprisingly, the evidence for the SM with quadratic divergences is minuscule compared to that for the SM without quadratic divergences. 
The Bayes-factors in \reftable{Table:SM_CMSSM_evidences} are more than $10^{30}$ against the SM with quadratic divergences ($150$ is considered \ic{very strong} on the Jeffreys' scale).

\begin{figure}[ht]
\centering
\includegraphics[height=0.49\linewidth]{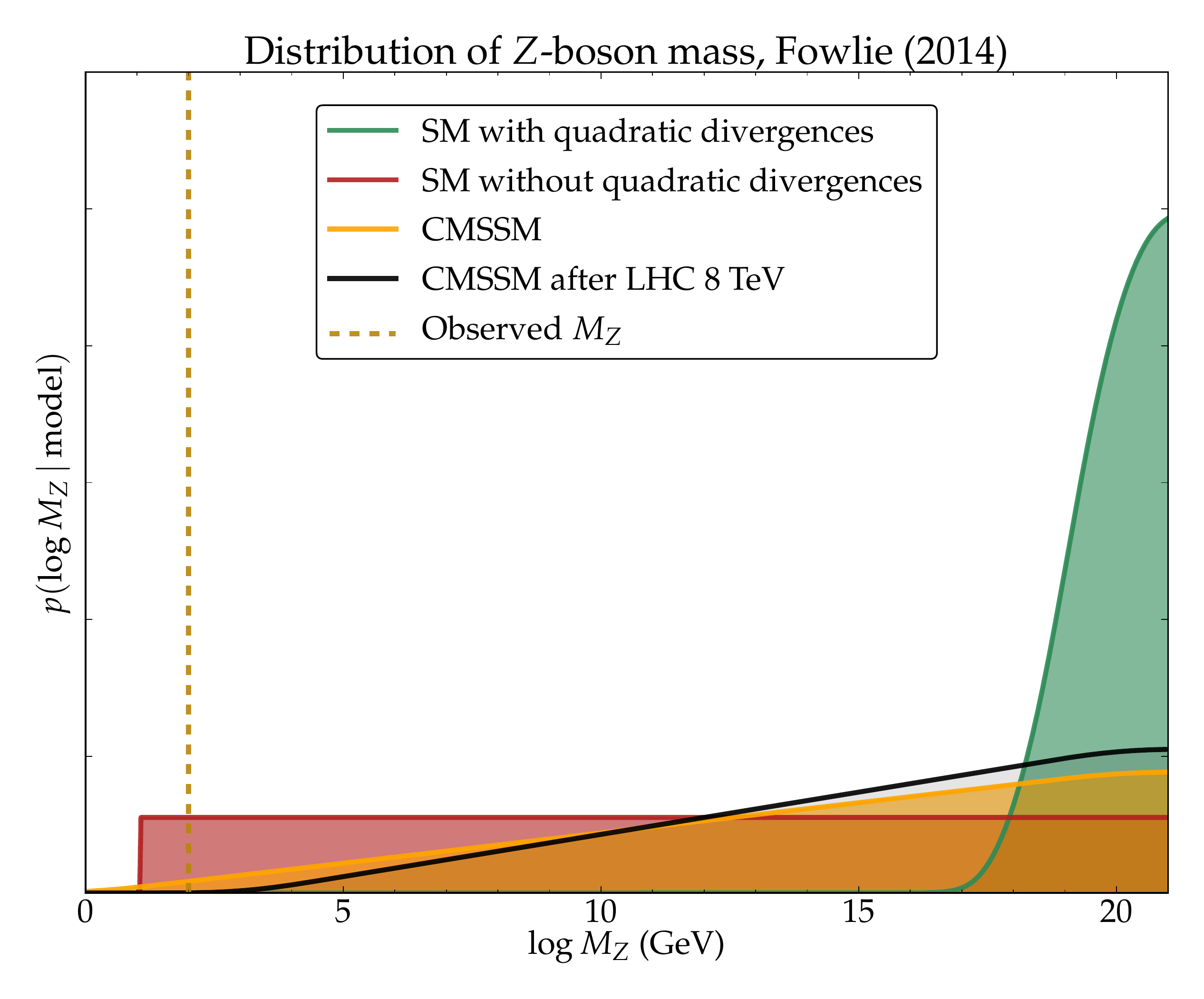}
\caption{The probability distribution of the $Z$-boson mass in the various models. The area under each plot is equal to one.}
\label{Fig:SM_Evidence}
\end{figure}

Let us interpret the evidence as a sampling distribution for the expected $Z$-boson mass, \ie plot the evidence as a function of \mz (\reffig{Fig:SM_Evidence}). As expected, the SM with quadratic divergences squanders its prediction for the $Z$-boson mass near \mplanck, far away from the measured \mz. The SM with quadratic divergences is unnatural. Because without quadratic divergences one can make no prediction for the magnitude of \mz, the SM without quadratic divergences is somewhat unnatural and complicated.

Now that we have completed the somewhat trivial exercise of calculating the evidences for the SM, let us calculate the evidences for the CMSSM.

\section{\label{Section:CMSSM_evidence}Bayesian evidences for the CMSSM}
The $Z$-boson mass, or, equivalently, the scale of electroweak symmetry breaking, is predicted in the MSSM via radiative electroweak symmetry breaking. At tree-level\cite{Martin:1997ns}, 
\beq
\label{Eqn:CMSSM_MZ}
\frac12 \mz^2 = - \mu^2 + \frac{\mhd^2 - \mhu^2 \tanbsq}{\tanbsq -1}.
\eeq
This expression is problematic; it contains the \ic{little-hierarchy problem}\cite{Barbieri:2000gf} and the related \ic{$\mu$-problem}\cite{Kim:1983dt}. From experiments, we know that \mz is $\sim 100\gev$. The MSSM predicts \mz via a cancellation between the SUSY breaking parameters, $\mhu^2$ and $\mhd^2$, and a SUSY preserving parameter in the superpotential, $\mu$. If the SUSY breaking scale is greater than the measured value of \mz, a cancellation between such large numbers is somewhat miraculous. This is the little-hierarchy problem. This problem is statistical in nature; we are concerned that the MSSM is unlikely because its parameters must be fine-tuned, \ie it might only agree with experiments in a small fraction of its parameter space. 

With a simplified \refeq{Eqn:CMSSM_MZ},
\beq
\frac12 \mz^2 \simeq -\mu^2 - \mhu^2,
\eeq
we found an analytic expression for the evidence from \refeq{Eqn:Evidence} as a function of \mz in the CMSSM with logarithmic priors. We plot this expression as a function of the $Z$-boson mass in \reffig{Fig:SM_Evidence}. Whilst the CMSSM is somewhat fine-tuned, the fine-tuning of the SM with quadratic divergences is far worse. The SM dimension-two coupling is quadratically sensitive to the UV; the highest scales must enter our expression for \mz. In the SM with quadratic divergences, the cancellation resulting in \mz must involve quantities $\sim\mplanck$.  In the CMSSM, we require a cancellation, but the cancellation could be at any scale up to \mplanck.  

Fine-tuning is typically measured with a sensitivity, for example, that originally proposed in \refcite{Ellis:1986yg,Barbieri:1987fn}, the Barbieri-Giudice measure,
\beq
\label{Eqn:BG}
\Delta_i = \frac{x_i}{\mz^2} \dpartial{\mz^2}{x_i},
\eeq
where $x_i$ are the model's parameters. The reciprocal of this measure is, indeed, similar to our Bayesian evidence, in that a small Barbieri-Giudice measure indicates that the model might spend its probability mass around the measured value of \mz. This is illuminated by rewriting \refeq{Eqn:BG},
\beq
\Delta_i^{-1} = \[\frac{\Delta\mz^2}{\mz^2}\frac{x_i}{\Delta x_i}\]^{-1} \propto \frac{\Delta x_i}{x_i}.
\eeq
The reciprocal of the Barbieri-Giudice measure is proportional to the local fraction of the model's parameter space in which \mz varies by $\Delta \mz$; in similarity, the evidence is a measure of the fraction of the model's prior volume in which the model agrees with experiments\cite{Cabrera:2008tj,Cabrera:2009dm,Fichet:2012sn}. The Barbieri-Giudice measure lacks, however, a formal interpretation and is, furthermore, a property of a point in the model's parameter space, rather than of the model itself (\latincf~the evidence).

Bayesian evidence automatically penalizes fine-tuning. Focusing mechanisms (\eg the focus-point\cite{Chan:1997bi,Feng:1999zg,Feng:2011aa}) are automatically incorporated. We must, however, choose \ic{honest} priors. In the CMSSM, we ought to formulate our prior beliefs in $\mu$ and $b$, the fundamental parameters, defined
\begin{align}
 W &\supset \mu H_u H_d,\\
 \mathcal{L}_{\text{Soft}} &\supset - b^2 H_u H_d + \text{c.c.}
\end{align}
For pragmatism, however, we exchange $\mu$ and $b$ for \mz and \tanb via \eg \refeq{Eqn:CMSSM_MZ}. We ought to transform our priors with the appropriate Jacobian, resulting in \textit{effective priors} for \mz and \tanb\cite{Allanach:2006jc,Allanach:2007qk,Cabrera:2008tj,Cabrera:2009dm,Cabrera:2012vu,Fichet:2012sn}.  With logarithmic priors for $\mu$ and $b$, our effective priors are
\begin{align}
\label{Eqn:BG_priors}
\priorf{\mz} &= \dpartial{\mu}{\mz} \priorf{\mu}  = \frac{2\mu}{\mz} \Delta_\mu^{-1} \priorf{\mu} = \text{const.}\,\Delta_\mu^{-1},\\
\priorf{\tanb} &= \dpartial{b}{\tanb} \priorf{b}  = \frac{\text{const.}}{b}\dpartial{b}{\tanb}.
\end{align}
The effective prior for \mz reveals the formal relationship between Bayesian statistics and the Barbieri-Giudice measure\cite{Cabrera:2008tj,Cabrera:2009dm}. With the Barbieri-Giudice measure, the statistical nature of the problem is latent\cite{Giusti:1998gz,Strumia:1999fr}; it is now manifest. 

We calculated the evidence exactly in the CMSSM with \ic{honest} priors. Let us make our prior choices clear, because it is a potential source of confusion. For the fundamental CMSSM parameters and priors we pick\footnote{One might wonder whether we should pick, \eg  $\mzero^2$ rather than \mzero as a fundamental parameter, since it is the square which appears in the soft-breaking Lagrangian. Because we pick logarithmic priors, however, the choice is irrelevant.}
\begin{align}
\mzero &\text{ log prior between } 1\gev \text{ and } \mplanck,\\\nonumber
\mhalf/\mzero &\text{ log prior between } 10^{-3} \text{ and } 10^3,\\\nonumber
\azero/\mzero &\text{ linear prior between } {-5} \text{ and } 5,\\\nonumber
b/\mzero &\text{ log prior between } 10^{-3} \text{ and } 10^3,\\\nonumber
\mu &\text{ log prior between } 1\gev \text{ and } \mplanck.
\end{align}
We anticipate that a breaking mechanism might distribute the SUSY breaking masses about a common SUSY breaking scale\cite{Cabrera:2009dm}, which we pick as \mzero. We do not consider mechanisms in which SUSY breaking parameters are split into distinct groups separated by many orders of magnitude\cite{Giudice:2004tc,ArkaniHamed:2004fb}. We call this choice of priors and parameterization our \textit{de jure} priors.

Were we to numerically calculate the evidence for the CMSSM with our \textit{de jure} priors, we would waste CPU time considering parameter space with incorrect \mz. For the purpose of our numerical calculation, we transform our \textit{de jure} priors into our equivalent \textit{de facto} priors,
\begin{align}
\mzero &\text{ log prior between } 1\gev \text{ and } 20\tev,\\\nonumber
\mhalf/\mzero &\text{ log prior between } 10^{-3} \text{ and } 10^3,\\\nonumber
\azero/\mzero &\text{ linear prior between } -5 \text{ and } 5,\\\nonumber
\tanb &\text{ effective prior between } 1 \text{ and } 60,\\\nonumber
\mz &\text{ effective prior, fixed } 91.1876\gev,
\end{align}
where the effective priors are in \refeq{Eqn:BG_priors}. The \ic{missing} parameter space in our \textit{de facto} priors at $\msusy\gg20\tev$ is irrelevant in our calculation, because it contains negligible evidence. The \ic{missing} parameter space, however, results in differences in normalization between our \textit{de jure} and \textit{de facto} priors, which we correct by hand. Fortunately, because the sign of $\mu$ is identical at the electroweak and \mplanck scales, we require no Jacobian to transform our prior for \sgnmu from \mplanck to the electroweak scale.  

We pick informative, Gaussian priors for the SM nuisance parameters \mts, \mbs, \invalphas and \alphass\cite{Beringer:1900zz}. \refcite{Cabrera:2008tj} stresses that the top and bottom masses are derived parameters; the input parameters are the Yukawa couplings, $y_t$ and $y_b$. In the CMSSM, the relationship between fermion masses and the Yukawa couplings includes factors of $\sin\beta$ and $\cos\beta$. We should pick priors for the Yukawa couplings rather than for the fermion masses; however, at leading order with logarithmic priors for the Yukawa couplings, there is no effective prior associated with $(y_t, y_b)\to(\mts,\mbs)$. At leading order, our treatment of the SM nuisance parameters is equivalent to picking logarithmic priors for the Yukawa couplings.

We calculated the CMSSM's mass spectrum and effective priors with \texttt{SOFTSUSY}\cite{Allanach:2001kg}. We used \texttt{MultiNest}\cite{Feroz:2008xx} with \texttt{PyMultiNest}\cite{Buchner:2014nha} to perform the integral in \refeq{Eqn:Evidence}. We found the evidence for three cases:
\begin{enumerate}
 \item $\mz=91.1876\gev$\cite{Beringer:1900zz} only in our likelihood (fixed by our \textit{de facto} priors),
 \item \mz, $\mh=125.9\pm0.4\pm2.0\gev$\cite{Beringer:1900zz,Chatrchyan:2012ufa,Aad:2012tfa,Allanach:2004rh} and the null result from the LHC in $20\invfb$\cite{ATLAS-CONF-2013-047} in our likelihood, and
 \item \mz, \mh and a hypothetical null result from the VLHC in $3000\invfb$\cite{Cohen:2013xda} in our likelihood.
\end{enumerate}
In the first case, our likelihood for \mz is a Dirac delta function. In the second case, our likelihood for \mh is a Gaussian with theoretical and experimental errors added in quadrature, and we veto points that are excluded by an ATLAS search for jets and missing energy\cite{ATLAS-CONF-2013-047}. In the last case, we consider the potential consequences of the \roots{100} VLHC, by vetoing points that would be excluded by a null result in $3000\invfb$\cite{Cohen:2013xda}, \ie points with $\ms{g}\lesssim16\tev$ and $\ms{q}\lesssim16\tev$ or points with $\ms{g}\lesssim13.5\tev$. 

\begin{table}[ht]
\begin{center}
\begin{ruledtabular}
\begin{tabular}{llll}
& \mz &  \mz, \mh and LHC & \mz, \mh and VLHC\\
 \hline
 Evidences, \ev & $\text{GeV}^{-1}$ & $\text{GeV}^{-2}$ & $\text{GeV}^{-2}$\\
\hline
SM with quadratic divergences & \pn{9}{-37} &  \pn{2}{-40}   & \pn{2}{-40}\\
SM no quadratic divergences                            & \pn{1}{-4}    &  \pn{2}{-7}     & \pn{2}{-7}\\
CMSSM						             & \pn{8}{-5}    &  \pn{3}{-10}  & \pn{7}{-13}\\
\hline
 Bayes-factors, $B=\ev_a/\ev_b$\\
\hline
CMSSM/SM with quadratic divergences      & \pn{9}{31} & \pn{2}{30} & \pn{4}{27}\\
SM no quadratic divergences/CMSSM                                & \pn{2}{0}    & \pn{7}{2} & \pn{3}{5}\\
\hline
 Partial Bayes-factors, $P=B_{i+1}/B_i$\\
\hline
SM no quadratic divergences/CMSSM & $\sim2$ & $\sim500$ & $\sim400$\\
%$\Delta\log_{10}$ SM $\Delta\mu^2=0$/CMSSM & $0.2$ & $2.7$ & $2.6$\\
\end{tabular}
\end{ruledtabular}
\end{center}
\caption{Bayesian evidences  and Bayes-factors for the SM with quadratic divergences, SM without quadratic divergences and CMSSM. The headings indicate which experimental results were included. The final column is the \ic{fine-tuning price,} as measured by partial Bayes-factors.}
\label{Table:SM_CMSSM_evidences}
\end{table}

\begin{figure}[ht]
\centering
\includegraphics[height=0.49\linewidth]{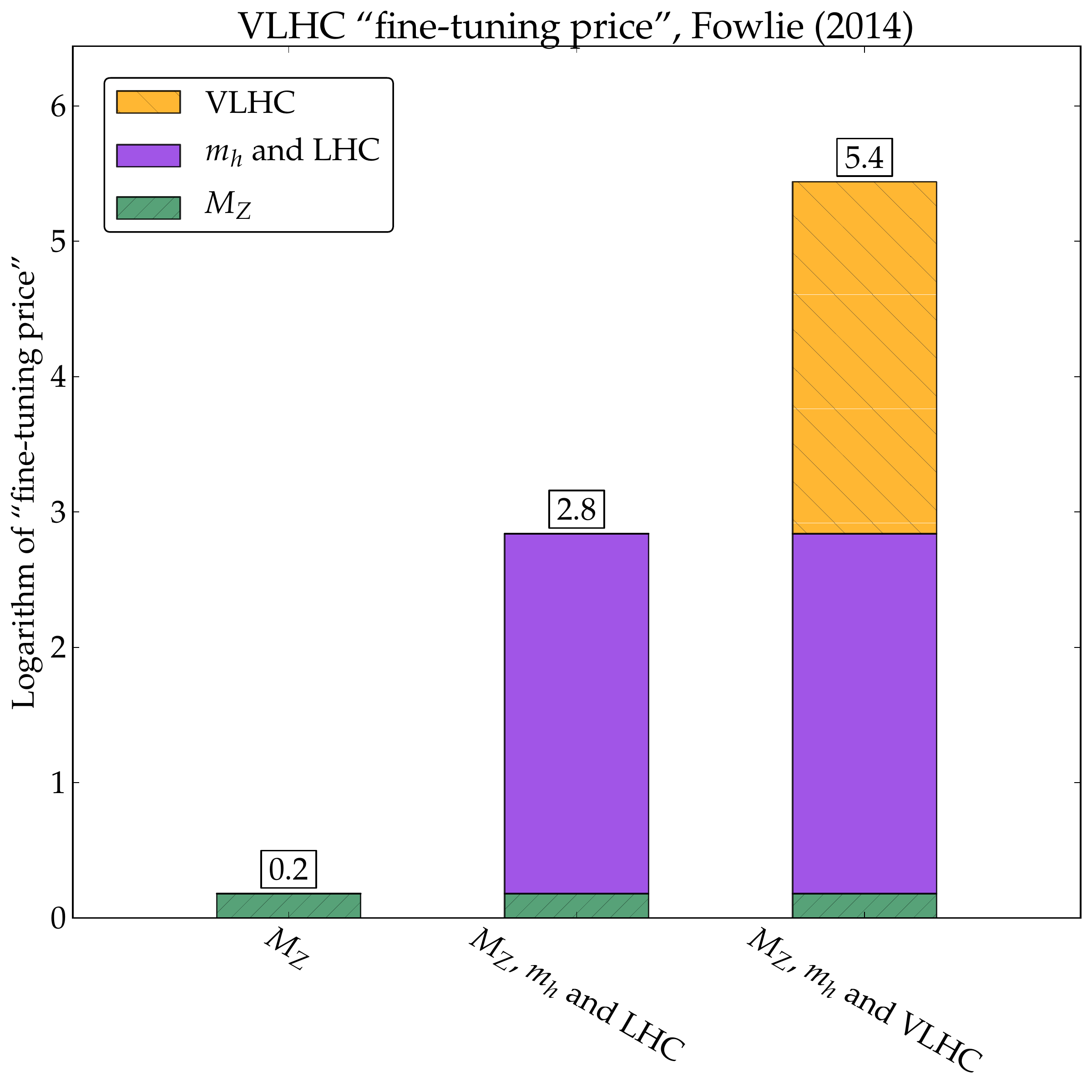}
\caption{The \ic{fine-tuning prices} of the \mz measurement, LHC experiments and hypothetical null results from the VLHC.  Our \ic{fine-tuning prices} are the Bayes-factors for the SM without quadratic divergences against the CMSSM broken down by experiment. \mz indicates the measurement of the $Z$-boson mass, \mh and LHC indicates the LHC Higgs mass measurement and null results from LHC, and VLHC indicates hypothetical  null results in $3000\invfb$ at \roots{100}. The logarithm is base 10.}
\label{Fig:FT_price}
\end{figure}

The evidences for the CMSSM in our three cases are shown are shown in \reftable{Table:SM_CMSSM_evidences}. Let us discuss the results case by case:
\begin{enumerate}
  \item\textit{\mz only in our likelihood.} The Bayes-factor favors the CMSSM over the SM with quadratic divergences by $\sim10^{32}$; as anticipated, the CMSSM is favored by naturalness. The Bayes-factor favors the SM without quadratic divergences over the CMSSM by only $\sim2$, which is \ic{barely worth mentioning} on the Jeffreys' scale in \reftable{Table:Jefferys_Scale}. Prior to LHC experiments, the CMSSM was not unnatural.
  
  \item\textit{\mz, \mh and LHC $20\invfb$ in our likelihood.} The Bayes-factor favors the CMSSM over the SM with quadratic divergences by $\sim10^{30}$; the little-hierarchy problem in the CMSSM is minuscule compared with the hierarchy problem in the SM with quadratic divergences.  The Bayes-factor favors the SM without quadratic divergences over the CMSSM by $\sim700$ ($150$ is \ic{very strong} on the Jeffreys' scale). Relative to the SM without quadratic divergences, the evidence for the CMSSM diminishes by a factor of $\sim500$; this is the \ic{fine-tuning price} of the LHC.
  
  \item\textit{\mz, \mh and a hypothetical null result from VLHC $3000\invfb$ in our likelihood.} The Bayes-factor favors the SM without quadratic divergences over the CMSSM by $\sim10^5$. Relative to the SM without quadratic divergences, the evidence for the CMSSM diminishes by a further factor of $\sim400$. The \ic{fine-tuning price} of null results from the VLHC ($\sim400$) would be similar to, though slightly less than that of the LHC ($\sim500$).

\end{enumerate}
Note that in all cases, however, the Bayes-factors favor the CMSSM over the SM with quadratic divergences by $\gtrsim10^{27}$. The \ic{fine-tuning prices} for the experiments in the CMSSM are illustrated in \reffig{Fig:FT_price} by the logarithm of the Bayes-factor for the SM without quadratic divergences against the CMSSM.  

The posterior probability density (see \eg \refcite{Fowlie:2011mb} for an introduction) is a by-product of the \texttt{MultiNest} evidence calculation.
With \mz, \mh and null results from the LHC in our likelihood, the posterior probability density  for \pmm confirms that the focus-point\cite{Chan:1997bi,Feng:1999zg,Feng:2011aa} at $\mzero\sim8\tev$ and $\mhalf\lesssim2\tev$ is favored. With only \mz in our likelihood,  unsurprisingly, we find that $\msusy\sim\mz$ is favored by \mz, \ie by naturalness.

\section{\label{Section:Fair_Comp}Effective versus UV-complete theories}
We interpreted the SM as an effective theory valid only below a cut-off scale, $\Lambda=\mplanck$, above which, we presumed, quantum field theory (QFT) is significantly modified by new physics (NP) related to gravity. The SM has a single relevant operator, $\mu^2\phi^2$. We considered the finite bare mass, $\mu^2$, to be a physical parameter originating from NP.  We parameterized our ignorance of $\mu^2$ with a logarithmic prior. We assumed that quadratic corrections to $\mu^2$ are unaffected by NP below the Planck scale, hinted at by \eg neutrino masses, inflation and dark matter.

If one attempts to remove the cut-off from the SM, $\Lambda\to\infty$, the bare mass diverges. The renormalized mass, $\mu_\text{R}^2$, might be considered to be fundamental. The renormalized mass differs from the bare mass by a quadratic correction and scheme-dependent terms. Because there are no quadratic corrections to the renormalized mass, it runs logarithmically from the Planck scale to the EW scale. The hierarchy problem is hidden in counter-terms. There are numerous problems with such an approach, \eg triviality.

We, however, interpreted the CMSSM as an ultra-violet (UV) complete theory valid at all scales. The fundamental parameters were renormalized SUSY breaking masses defined at the renormalization scale $\mu=\mgut$ in the minimal subtraction scheme, rather than bare masses. Was it fair to compare the SM as an effective theory with the CMSSM as a UV-complete theory? Whilst with Bayesian evidence one can compare any models that make predictions for the experimental data, comparisons are interesting only if the models are realistically interpreted.

% Suppose we interpret the CMSSM as an effective theory valid only below a cut-off scale, $\Lambda=\mgut$, at which new GUT physics is important. Divergences are no worse than logarithmic in supersymmetric models. In the minimal subtraction scheme, the renormalized SUSY breaking masses would equal the bare SUSY breaking masses at the GUT scale. Suppose we instead interpret the CMSSM as an effective theory valid only below the Planck scale at which SUSY breaking is mediated to the CMSSM from a hidden sector via gravitational interactions. The CMSSM is defined by bare SUSY breaking masses unified at the Planck scale. The renormalized masses at the GUT scale would be related to the unified bare masses at the Planck scale by
% \beq
% m_\text{R}^2\(\mu=\mgut\) = m^2\(\Lambda=\mplanck\) \[1+\frac{\alpha}{\pi}\ln\(\frac{\mplanck}{\mgut}\)\],
% \eeq
% where scheme dependent corrections are omitted and $\alpha$ is a typical coupling strength dependent on the new GUT physics\cite{Polonsky:1994sr}. These differences are likely to be small (though could spoil unification at the GUT scale). 

Suppose we instead interpreted the CMSSM as an effective theory valid only below a cut-off scale, $\Lambda=\mgut$, at which new GUT physics is important, or the Planck scale, at which gravitational interactions mediate SUSY breaking.  Divergences are no worse than logarithmic in supersymmetric models. The bare SUSY breaking masses at the cut-off scale would be similar to renormalized SUSY breaking masses at the GUT or Planck scales; the bare and renormalized masses would differ by logarithmic corrections. No fine-tuning of the EW scale is hidden by parameterizing the CMSSM as a UV-complete theory with renormalized masses.\footnote{It is possible, however, that focusing mechanisms are disfavored if SUSY breaking masses are unified at the Planck scale rather than at the GUT scale\cite{Polonsky:1994sr}.}

The comparison was fair, though its outcome was perhaps inevitable. Although the SM was vastly disfavored, it was important in the analysis; it was a reference model against which we judged the severity of the change in the \ic{fine-tuning price} in the CMSSM.

\section{\label{Section:Mu_problem}The \texorpdfstring{$\mu$}{mu}-problem}
A problem emerges from our \ic{honest} choice of prior for $\mu$, which aggravates the fine-tuning problem. The $\mu$-parameter is a symmetry conserving parameter in the superpotential. \textit{A priori,} it is unrelated to a symmetry breaking scale. This is problematic; phenomenologically it must be that $\mu \sim \msusy$. 
The evidence for a model in which we expect $100\gev\lesssim\mu\lesssim\mplanck$ and observe $\mu \sim \msusy$ could be  smaller than that for a model in which we expect $\mu \sim \msusy$ and observe $\mu \sim \msusy$. 
This is the \ic{$\mu$-problem;} in our formulation, its statistical nature is manifest.  \refeq{Eqn:BG_priors} reveals the $\mu$-problem and the fine-tuning problem;  the $\mu$-problem is that \priorf{\mu\approx\msusy} is small and the fine-tuning problem is that \indpartial{\mu}{\mz} is small, resulting in a small prior belief in the observed electroweak scale, \priorf{\mz}. The ratio of evidences for a model that predicts $\mz\lesssim\mu\lesssim\mplanck$ and an \ic{almost-so} model that predicts \eg $10^{-1}\msusy\lesssim\mu\lesssim10^3\msusy$ is approximately
\beq
\frac{\ln\(\frac{10^{3}\msusy}{10^{-1}\msusy}\)}{\ln\(\frac{\mplanck}{\mz}\)}\approx \frac15.
\eeq
A similar result applies to SUSY models with a Giudice-Masiero mechanism\cite{Giudice:1988yz}.

The little-hierarchy problem in the CMSSM is $\sim30$ times worse than the $\mu$-problem. The $\mu$-problem contributes a factor of only $\sim5$ to a Bayes-factor for an \ic{almost-so} model against the CMSSM. The Bayes-factor with \mz, \mh and null results from the LHC favors the SM without quadratic divergences over the CMSSM by $\sim700$; the little-hierarchy problem contributes a factor of  $\sim150$ and the $\mu$-problem contributes a factor of $\sim5$.

\section{\label{Section:Conclusions}Conclusions}
The absence of SUSY or other new physics at the LHC has lead many to question naturalness arguments. Drawing upon the literature, we clarified the relationship between Bayesian statistics and naturalness, concluding that natural models are most probable and that naturalness is not merely an aesthetic principle. We calculated the Bayesian, probabilistic measure of naturalness, the evidence, for the SM with and without quadratic divergences, demonstrating that the SM with quadratic divergences is improbable. We calculated the evidence for the CMSSM in three cases: with only the \mz measurement; with the \mz measurement and LHC measurements; and with the \mz measurement and a hypothetical null result from the VLHC with $3000\invfb$. 
The latter allowed us to quantitatively understand the potential \ic{fine-tuning price} of the VLHC. We found that the \ic{fine-tuning price} of null results from the VLHC ($\sim400$) would be slightly less than that of the LHC  ($\sim500$).  We hope this result might help to inform preliminary discussions and plans for the VLHC.

\begin{acknowledgements}
I thank M.~Raidal and A.~Strumia for helpful criticisms of my manuscript.
This work was supported in part by grants IUT23-6, CERN+,  and by the European Union
through the European Regional Development Fund and by ERDF project 3.2.0304.11-0313
Estonian Scientific Computing Infrastructure (ETAIS).
\end{acknowledgements}

\let\cite\oldcite
\bibliography{main}
\end{document}